\documentclass[letterpaper, 10 pt, conference]{ieeeconf}  

\IEEEoverridecommandlockouts                              

\usepackage{amssymb}
\usepackage{amsmath}

\usepackage{graphics} 
\usepackage{epsfig} 
\usepackage{mathptmx} 
\usepackage{times} 
\usepackage{booktabs}
\usepackage{multirow} 
\usepackage{array}    
\usepackage{siunitx}  
\usepackage{framed}        
\usepackage{multicol}
\usepackage{nomencl}
\usepackage{algorithm}
\usepackage[noend]{algpseudocode}
\makeatletter
\renewcommand\fs@ruled{\def\@fs@cfont{\bfseries}\let\@fs@capt\floatc@ruled
  \def\@fs@pre{\vskip6pt\hrule height.8pt depth0pt \kern2pt}%
  \def\@fs@post{\kern2pt\hrule\relax}%
  \def\@fs@mid{\kern2pt\hrule\kern2pt}%
  \let\@fs@iftopcapt\iftrue}
\makeatother

\usepackage{subcaption}

\usepackage{xurl}   
\usepackage{placeins}
\usepackage{graphicx}

\newtheorem{remark}{Remark}

\newtheorem{proposition}{Proposition}

\newcommand{\Pb}{P^{\mathrm{buy}}}
\newcommand{\Ps}{P^{\mathrm{sell}}}
\newcommand{\Gb}{G^{\mathrm{buy}}}
\newcommand{\Gs}{G^{\mathrm{sell}}}
\newcommand{\Gt}{G^{\mathrm{tot}}}
\newcommand{\Ct}{C^{\mathrm{tot}}}
\newcommand{\Cb}{C^{\mathrm{buy}}}
\newcommand{\Cs}{C^{\mathrm{sell}}}
\newcommand{\cC}{\mathcal{C}}
\newcommand{\SoC}{\mathrm{SoC}}
\newcommand{\SoCmax}{\overline{\mathrm{SoC}}}
\newcommand{\cA}{\mathcal{A}}
\newcommand{\Anew}{\mathcal{A}_{\mathrm{new}}}
\newcommand{\st}{\tilde s}

\title{\LARGE \bf
Resilient Coalition Formation with Proactive Scheduling and Outage Management for Energy Management 
}

\author{Haochen Tao$^{1}$, Luke Rickard$^{2}$ and Francesca Boem$^{1}$
\thanks{*This work was supported by the EPSRC grant EP/Z536106/1. ``Control Methods for Cyber-Physical Resilience of AI-enabled Critical Infrastructure Systems'' (LTRF-2526-22-215)
was supported by The Leverhulme Trust (ROR ID: https://ror.org/012mzw131) and the Royal Academy of Engineering (https://ror.org/0526snb40) under the RAEng/Leverhulme Trust Research Fellowships scheme.}
\thanks{$^{1}$Department of Electronic and Electrical Engineering, University College London, London, UK. $^{2}$University of East London, London, UK.}
}

\begin{document}

\maketitle
\thispagestyle{empty}
\pagestyle{empty}

\begin{abstract}
As extreme weather events that affect the power grid become more frequent, keeping buildings supplied during outages of the upstream grid has become a critical challenge.
This work proposes a resilient coalitional control scheme, in which buildings with local generation and batteries form coalitions that share energy among their members.
The coalitions are formed to maximize the resilience of the buildings, so that the least resilient coalition, with the smallest margin of stored energy over its reserve, is as resilient as possible.
Each coalition adopts a Model Predictive Control-based approach for proactive scheduling and outage management.
Proactive scheduling minimizes the operation cost while letting each coalition keep a reserve of stored energy to prepare for a possible outage. Outage management minimizes load shedding and generation curtailment in the event of a grid fault.
Simulations on $12$ buildings show that the proposed method reduces both the load shedding and the generation curtailment in the case of outages, compared with decentralized control.

\end{abstract}

\section{INTRODUCTION}

Climate change is increasing the frequency and severity of several extreme weather events~\cite{IPCC2021Ch11}, which are already having adverse effects on energy distribution networks~\cite{PANTELI2015259}. 
Buildings are the majority of participants in energy distribution networks, and are frequently being adapted with storage and generation capabilities, such buildings may be termed prosumers (participants in the energy market who both consume and produce electricity~\cite{FILHO2024100158,Parag2016}).
When an extreme event disconnects the upstream grid, their local generation and storage can supply their own demand.
Here, resilience is the ability of the buildings to keep supplying their demand during an outage.

We consider the problem of resilience against grid outages, where buildings are prevented from buying from or selling to the grid. 
Two strategies are commonly proposed to increase the resilience to such faults, namely proactive scheduling and outage management~\cite{HUSSAIN201956,casagrande2022resilient,tao2025resilient}.
Proactive scheduling allows a building to store a reserve of energy in preparation for a grid fault of a given length.
For a single microgrid, resilience-oriented scheduling with robust or stochastic optimization for the uncertain duration of the outage is applied in~\cite{zografou2020microgrid,vahedipour2020risk}.
On the other hand, outage management considers how a building should behave during an outage to reduce load shedding and generation curtailment whilst unable to trade with the grid.
The scheduling requirement is enforced as a chance constraint of a stochastic MPC, taking into account the fault probability and with forecasts and their uncertainty learned online in~\cite{casagrande2024learning}, while proactive scheduling and outage management are proposed in the same framework with a convex optimization approach in~\cite{tao2025resilient}. However, both works consider a single microgrid with a central controller.
During an outage, a distribution network can also be sectionalized into self-supplied microgrids~\cite{shimillas2025adaptive}.
Since the resources of a single building or microgrid are limited, networked microgrids may share their generation and storage to ride through extreme events~\cite{bordbari2024networked}, for instance with a hierarchical outage management scheme for multi-microgrids~\cite{mansouri2022hierarchical}.
However, a building that acts alone may be unable to hold enough stored energy for an outage.

Rather than fully decentralized operation, the buildings can form small groups that share energy and information only among their members.
Coalitional control~\cite{7823093,DBLP:conf/amcc/FeleMC15} is an expanding area of research which considers how cooperative agents may form into disjoint groups for mutual benefits. 
Unlike work on distributed optimization~\cite{boyd2011admm,nedic2009distributed,casagrande2022resilient}, we do not wish for agents to work in a single unified group (sometimes referred to as the grand coalition), in order to reduce communication overheads and unnecessary information sharing~\cite{DBLP:journals/corr/abs-2603-28562}.
In energy systems, coalitional game theory has been used to form coalitions of microgrids that reduce power losses and energy costs~\cite{modarresi2022coalitional}, and provide resilience against extreme events~\cite{gautam2023graph}.
However, these works utilizing coalitional games do not consider proactive steps to prepare for possible later outages or faults and specific management in faulty conditions, whereas coalitional control forms the coalitions in a closed loop as the system evolves.
In our framework, coalitions are groups of buildings that may trade energy among themselves, improving resilience, and reducing costs.

We directly consider resilience in the dynamic coalition formation process; we address both proactive scheduling and outage management problems, make use of coalitions to reduce the storage requirements on individual buildings for proactive scheduling, and improve performance during outage management.
Since proactive scheduling requires a lookahead into the future, we adopt  Model Predictive Control (MPC) techniques to optimize energy trades and battery charging profiles, with forecasted energy demand and generation over the prediction horizon. 
In the single building energy management context, this approach has been studied in~\cite{oldeWurtel2012mpc}, whereas we focus on multi-building control.
Our work extends upon current applications of MPC to coalitions~\cite{baldivieso2021coalitional,sanchez2023robust} by considering resilience as the key driver for coalition formation, as opposed to economic benefit.

We propose a coalition formation technique which addresses the problem of resilience to grid faults.
In particular, we use a resilience metric to group buildings in a way that increases the resilience of all individual buildings, both during proactive scheduling and outage management.
The main contributions of this paper are as follows:

\begin{itemize}
\item We propose a novel coalition formation metric that seeks to increase resilience of individual buildings;
\item We consider coalition formation both for proactive scheduling, allowing buildings to pool energy storage, and for outage management, allowing buildings to trade energy without a grid connection, reducing load shedding and generation curtailment;
\item  Numerical results demonstrate improved resilience over decentralized control.
\end{itemize}

\section{Problem formulation}\label{sec:model}

This section presents the models of the considered buildings, their coalitions and connection to the upstream grid. 

\subsection{Buildings Models}
The system consists of $N$ buildings denoted as $\mathcal{M}_1,\dots,\mathcal{M}_N$, operated over the discrete time steps $t\in\mathcal T=\{1,\dots,T\}$ with a sampling time of $T_s$ minutes.
At each time step, each building $\mathcal{M}_i$ has a local aggregated energy demand $d_i(t)\geq0$ and local generation $g_i(t)\geq0$ from renewable sources. Here $d_i(t)$, $g_i(t)$ and all the other energy flows in this paper are energies over one time step. Each building is also equipped with a battery with capacity $\SoCmax_i$, charging and discharging efficiencies $\eta^c_i,\eta^d_i\in(0,1)$, and a maximum energy $\bar u_i$ that can be charged or discharged in one step.
The stored energy $\SoC_i(t)\in[0,\SoCmax_i]$ of the battery evolves as
\begin{equation}\label{eq:m_dyn}
\SoC_i(t+1)=\SoC_i(t)+\eta^c_iu^c_i(t)-\tfrac{1}{\eta^d_i}u^d_i(t),
\end{equation}
where $u^c_i(t)$ and $u^d_i(t)$ are the energies charged into and discharged from the battery during step $t$.

During normal operation, each building can buy energy $\Gb_i(t)$ from the upstream grid at the price $\Pb(t)$ and sell energy $\Gs_i(t)$ at the price $\Ps(t)$, with $\Pb(t)>\Ps(t)>0$.
The energy balance of building $\mathcal{M}_i$ is
\begin{equation}\label{eq:m_power0}
d_i(t)+u^c_i(t)=g_i(t)+u^d_i(t)+\Gt_i(t),
\end{equation}
where $\Gt_i=\Gb_i-\Gs_i$ is the net energy bought from the grid, and the traded energies are nonnegative
\begin{equation}\label{eq:m_pos0}
\Gb_i(t),\ \Gs_i(t)\ge0.
\end{equation}
During an outage period $\mathcal T^{\mathrm{out}}$, the connection to the upstream grid is lost, so $\Gt_i(t)=\Gb_i(t)=\Gs_i(t)=0$ in~\eqref{eq:m_power0} and the demand can be supplied only by local generation and storage.
The load part that cannot be supplied is shed, and the generation that can be neither used nor stored is curtailed.

\subsection{Coalitions and Energy Sharing}
\begin{figure}[tb]
\vspace{4pt}
\centering
\includegraphics[width=\columnwidth]{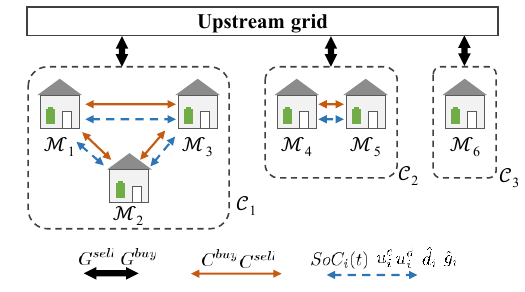}
\caption{Example of a coalition structure. Six buildings in three coalitions, with the energy exchange within a coalition represented by solid orange arrows, the energy exchanged with the grid by solid black arrows and the shared information by blue dashed arrows.}
\label{fig:coal_example}
\end{figure}

At each time step the buildings are divided into $M$ disjoint coalitions $\cC_j$, $j=1,\dots,M$, where $M$ and the coalitions can change from one step to the next, and every building belongs to exactly one coalition.
Such a division is called a coalition structure.
Each coalition is a set of at most $\mathcal{C}_{\max}$ buildings, and $i\in\cC_j$ denotes that $\mathcal{M}_i$ is a member of $\cC_j$.
The members within a coalition can exchange energy with each other, where $\Cb_i$ and $\Cs_i$ are the energy bought from and sold to the other members.
Thus $u$, $G$ and $C$ denote the energies exchanged with the battery, with the grid and within the coalition, respectively.
The energy balance~\eqref{eq:m_power0} of a building in a coalition then becomes
\begin{equation}\label{eq:m_power}
d_i(t)+u^c_i(t)=g_i(t)+u^d_i(t)+\Gt_i(t)+\Ct_i(t),
\end{equation}
where $\Ct_i(t)=\Cb_i(t)-\Cs_i(t)$ is the net energy bought from the other members, and the following constraints must be satisfied for the energy to be balanced within the coalition:
\begin{align}
\Cb_i(t),\ \Cs_i(t)&\ge0, \label{eq:m_pos}\\
\textstyle\sum_{i\in \cC_j}\Ct_i(t)&=0. \label{eq:m_coal}
\end{align}
Fig.~\ref{fig:coal_example} illustrates an example of a coalition structure.

\section{RESILIENT COALITIONAL CONTROL DESIGN}\label{sec:mpc}

Building on the model of Section~\ref{sec:model}, this section designs the resilient coalitional control, allocating at each step buildings to coalitions and scheduling their operations. 
The proposed framework combines an MPC-like approach for proactive scheduling and outage management, with a coalition formation based on a novel resilience metric.

\subsection{Proactive Scheduling}\label{sec:gridprob}
During normal operation, to be resilient to potential outages, each coalition employs proactive scheduling that keeps enough stored energy to supply its forecasted net demand over a window of future steps, while minimizing energy cost.
We introduce the predicted demand $\hat d_i(t+k)$ and generation $\hat g_i(t+k)$, which are the forecasts of $d_i(t+k)$ and $g_i(t+k)$ computed at time $t$.
The forecasts of the prices are denoted as $\hat P^{\mathrm{buy}}(t+k)$ and $\hat P^{\mathrm{sell}}(t+k)$.
We assume that these are provided by a suitable predictor, as in~\cite{casagrande2024online}; the design of the predictor is out of the scope of this paper.
At each time $t$, the proactive scheduling defines the reserve $\st_{\cC}$ as a lower bound on the total stored energy a coalition will hold for the next $\tau$ steps
\begin{equation}\label{eq:stilde}
\begin{split}
\st_{\cC}(t+k)=\max\Big\{0,&\ \min\Big\{\sum_{i\in \cC}\sum_{l=1}^{\tau}\big(\hat d_i(t+k+l)\\
&-\hat g_i(t+k+l)\big),\ \sum_{i\in \cC}\SoCmax_i\Big\}\Big\},
\end{split}
\end{equation}
where $\tau$ is the length of the proactive scheduling window.
The double sum is the predicted net demand of $\cC$ over the $\tau$ steps after $t+k$, hence it prepares the coalition to supply its predicted net demand for $\tau$ steps of outage.
The reserve is limited to the total capacity of the batteries of $\cC$, which cannot store more energy, and it is zero when the predicted generation exceeds the predicted demand.
Equation~\eqref{eq:stilde} is defined for the coalition as a whole, hence in a coalition with multiple members, a member may hold less than its own required reserve to satisfy local demand, when the other members of its coalition hold more.
At the beginning of each step $t$ before the coalitions are formed, the stored energy $\SoC_{i,t}$, the demand $d_i(t)$ and the generation $g_i(t)$ are measured, hence $\hat d_i(t)=d_i(t)$ and $\hat g_i(t)=g_i(t)$.
The members of a coalition share with each other their demand and generation $\hat d_i(t+k)$ and $\hat g_i(t+k)$ for $k=0,\dots,K+\tau-1$, where $K$ is the prediction horizon, and their stored energy $\SoC_{i,t}$, so that the operation of each coalition is planned jointly over the horizon.
The optimization problem for energy management with proactive scheduling in coalition $\cC_j$ at time $t$ is formulated as
\begin{subequations}\label{eq:P}
\begin{align}
&V(\cC_j)=\min_{x^{\mathrm{p}}}\ \sum_{i\in \cC_j}\sum_{k=0}^{K-1}\hat P^{\mathrm{buy}}(t+k)\,\Gb_i(t+k) \notag\\
&\hspace{9em}-\hat P^{\mathrm{sell}}(t+k)\,\Gs_i(t+k) \label{eq:obj}\\
&\text{s.t.}\ \hat d_i(t+k)+u^c_i(t+k)=\hat g_i(t+k)+u^d_i(t+k) \notag\\
&\hspace{6.5em}+\Gt_i(t+k)+\Ct_i(t+k) \label{eq:power}\\
&\phantom{\text{s.t.}\ }\SoC_i(t+k+1)=\SoC_i(t+k)+\eta^c_iu^c_i(t+k) \notag\\
&\hspace{6.5em}-\tfrac{1}{\eta^d_i}u^d_i(t+k) \label{eq:dyn}\\
&\phantom{\text{s.t.}\ }\SoC_i(t) = \SoC_{i,t} \label{eq:init}\\
&\phantom{\text{s.t.}\ }\SoC_i(t+k)\in[0,\SoCmax_i],\ k=1,\dots,K \label{eq:soc_con}\\
&\phantom{\text{s.t.}\ }u^c_i(t+k),\ u^d_i(t+k)\in[0,\bar u_i] \label{eq:control_con}\\
&\phantom{\text{s.t.}\ }\Cb_i(t+k),\ \Cs_i(t+k)\ge0 \label{eq:pos}\\
&\phantom{\text{s.t.}\ }\Ct_i(t+k) = \Cb_i(t+k)- \Cs_i(t+k) \label{eq:c_sum}\\
&\phantom{\text{s.t.}\ }\textstyle\sum_{i\in \cC_j}\Ct_i(t+k)=0 \label{eq:coal}\\
&\phantom{\text{s.t.}\ }\Gb_i(t+k),\ \Gs_i(t+k) \ge 0 \label{eq:gridpow}\\
&\phantom{\text{s.t.}\ }\Gt_i(t+k)=\Gb_i(t+k)- \Gs_i(t+k) \label{eq:gridtot}\\
&\phantom{\text{s.t.}\ }\textstyle\sum_{i\in \cC_j}\SoC_i(t+k)\ \ge\ \st_{\cC_j}(t+k), \notag\\
&\hspace{6.5em}k=1,\dots,K-1 \label{eq:floor}
\end{align}
\end{subequations}
for all $i\in \cC_j$ and for $k=0,\dots,K-1$ unless specified otherwise. $x^{\mathrm{p}}$ denotes the decision variables of~\eqref{eq:P}, as 
\begin{multline*}
x^{\mathrm{p}}\triangleq\big\{u^c_i(t+k),\ u^d_i(t+k),\\
\Cb_i(t+k),\ \Cs_i(t+k)\big\}_{i\in \cC_j,\ k=0,\dots,K-1}.
\end{multline*}
The objective~\eqref{eq:obj} is the cost of the energy bought from the grid minus the revenue of the energy sold over the horizon.
Constraint~\eqref{eq:floor} keeps the total stored energy of the coalition above the reserve~\eqref{eq:stilde} at every future step of the horizon.

\subsection{Outage Management}\label{sec:outage}
In the case of a grid outage, during $t\in\mathcal T^{\mathrm{out}}\subseteq\mathcal T$, the connection to the upstream grid is lost. To minimize load shedding and generation curtailment during faulty condition, each coalition solves the following optimization problem:
\begin{subequations}\label{eq:O}
\begin{align}
&V^{\mathrm{out}}(\cC_j)=\min_{x^{\mathrm{out}}}\ \sum_{i\in \cC_j}\sum_{k=0}^{K-1}c^{\mathrm{shed}}d^{\mathrm{shed}}_i(t+k) \notag\\
&\hspace{4em}+c^{\mathrm{curt}}g^{\mathrm{curt}}_i(t+k)+\varepsilon\,u^c_i(t+k) \label{eq:Oobj}\\
&\text{s.t.}\ \hat d_i(t+k)-d^{\mathrm{shed}}_i(t+k)+u^c_i(t+k) \notag\\
&\qquad=\hat g_i(t+k)-g^{\mathrm{curt}}_i(t+k) \notag\\
&\qquad\phantom{=}+u^d_i(t+k)+\Ct_i(t+k) \label{eq:Opower}\\
&\phantom{\text{s.t.}\ }0\le d^{\mathrm{shed}}_i(t+k)\le\hat d_i(t+k), \notag\\
&\phantom{\text{s.t.}\ }0\le g^{\mathrm{curt}}_i(t+k)\le\hat g_i(t+k) \label{eq:Obox}\\
&\phantom{\text{s.t.}\ }\text{\eqref{eq:dyn}--\eqref{eq:coal}} \notag
\end{align}
\end{subequations}
for all $i\in \cC_j$ and $k=0,\dots,K-1$. $d^{\mathrm{shed}}_i$ is the load shed, $g^{\mathrm{curt}}_i$ the generation curtailed, and $c^{\mathrm{shed}}$ and $c^{\mathrm{curt}}$ their cost coefficients, $\varepsilon>0$ is the penalty coefficient.
The plan assumes no grid connection over the whole MPC horizon, hence $\Gt_i=0$ and there is no reserve constraint~\eqref{eq:floor}.
$x^{\mathrm{out}}$ denotes the decision variables for~\eqref{eq:O}:
\begin{multline*}
x^{\mathrm{out}}\triangleq\big\{u^c_i(t+k),\ u^d_i(t+k),\ \Cb_i(t+k),\ \Cs_i(t+k)\\\big\}_{i\in \cC_j,\ k=0,\dots,K-1}.
\end{multline*}

\begin{proposition}[No Simultaneous Charge/Discharge]\label{prop:nosimO} \hfill
\begin{enumerate}[label=\roman*)]
    \item Every optimal solution of~\eqref{eq:P} satisfies $u^c_i(t+k)\,u^d_i(t+k)=0$ for all $i\in\cC_j$ and $k=0,\dots,K-1$.
    \item If $\varepsilon>c^{\mathrm{curt}}\max_{i=1,\dots,N}(1-\eta^c_i\eta^d_i)$, then at every step $k$ at which a member of $\cC_j$ sheds load or has generation that is not curtailed, every optimal solution of~\eqref{eq:O} satisfies $u^c_i(t+k)\,u^d_i(t+k)=0$ for all $i\in\cC_j$.
\end{enumerate}
\end{proposition}
The proof is detailed in the appendix.
Proposition~\ref{prop:nosimO} excludes simultaneous charging and discharging without the constraint $u^c_i(t+k)\,u^d_i(t+k)=0$, which would require binary variables and turn~\eqref{eq:P} and~\eqref{eq:O} into mixed-integer programs.
Such programs are NP-hard in general, while the coalition formation solves~\eqref{eq:P} or~\eqref{eq:O} for several candidate coalitions at every step.

\subsection{Resilient Coalition Formation}\label{sec:resilience}

This subsection presents the proposed coalition formation process at each time step and its objective, and it defines a resilience metric to compare coalitions.

\subsubsection{Resilience metric}
We define the reserve margin of a coalition $\cC$ as
\begin{equation}\label{eq:m}
m(\cC,t)=\sum_{i\in \cC}\big( g_i(t)+\SoC_i(t)-d_i(t)\big)-\st_{\cC}(t),
\end{equation}
which is the residual energy that the buildings in $\cC$ jointly have at step $t$ after supplying their net demand and removing the proactive scheduling reserve $\st_{\cC}(t)$.
While $m(\cC,t)$ only considers the current step,~\eqref{eq:floor} must hold at every later step of the horizon.
Since each battery stores at most $\eta^c_i\bar u_i$ at every step, the reserve requirement could in principle grow faster than the batteries can be charged, and a coalition with $m(\cC,t)\ge0$ may then be unable to satisfy~\eqref{eq:floor}, resulting in an infeasible scheduling problem. 
We avoid this infeasibility issue by only merging coalitions that allow feasible solutions.
To do this, the minimum reachable margin of a coalition $\cC$ is then defined as
\begin{multline}\label{eq:r}
r(\cC,t)=\min_{k=1,\dots,K-1}\Big[\sum_{i\in \cC}\min\big\{\SoCmax_i,\\
\SoC_i(t)+\eta^c_i\bar u_i k\big\}-\st_{\cC}(t+k)\Big],
\end{multline}
which is the difference between the maximum energy the members can store by charging $\bar u_i$ at each step and the reserve required, measured at the worst step in the horizon.
During normal operation, Problem~\eqref{eq:P} for $\cC$ is feasible if and only if $r(\cC,t)\ge0$, since no plan stores more energy than charging every battery at full power until it is full, and this plan satisfies all the other constraints of~\eqref{eq:P}.

The two margins are combined into the resilience value
\begin{equation}\label{eq:mu}
\mu(\cC,t)=\begin{cases}\min\{m(\cC,t),\,r(\cC,t)\}, & r(\cC,t)\ge0,\\ -\infty, & r(\cC,t)<0.\end{cases}
\end{equation}
Thus $\mu(\cC,t)\ge0$ if and only if $m(\cC,t)\ge0$ and $r(\cC,t)\ge0$, i.e., the coalition covers its net demand at step $t$ while keeping the reserve, and can reach the reserve over the whole MPC horizon.
If $r(\cC,t)<0$, then~\eqref{eq:P} is infeasible for the coalition $\cC$, and $\mu(\cC,t)=-\infty$ ranks $\cC$ below every feasible coalition.
This occurs, for example, if a member that holds less than its own reserve in a coalition is left alone at a later time step.
In order to avoid infeasibility, we propose to form the coalitions so that the weakest coalition is as resilient as possible, thus we seek to solve
\begin{equation}\label{eq:maxmin}
\max_{\{\cC_j\}_{j=1}^{M}}\ \min_{j=1,\dots,M}\ \mu(\cC_j,t),
\end{equation}
where the maximum is taken over all the possible coalition structures $\{\cC_j\}_{j=1}^{M}$ with $|\cC_j|\le \mathcal{C}_{\max}$, and $M$ depends on the structure.
At each time $t$, the coalition structure is selected by~\eqref{eq:maxmin}. In the case that more than one structure has the same value of~\eqref{eq:maxmin}, we select the one with the lowest total cost $\sum_{j}V(\cC_j)$, or $\sum_{j}V^{\mathrm{out}}(\cC_j)$ during an outage.
Each coalition then solves Problem~\eqref{eq:P}, or~\eqref{eq:O} during an outage, and applies the first step of its optimal plan.

\subsubsection{The algorithm}\label{sec:formation}

We now formalize the proposed methodology as an algorithm.
We introduce Algorithm~\ref{alg:form}, which illustrates how to compute the coalitions at each time step.
Since the number of coalition structures grows rapidly with $N$, the problem~\eqref{eq:maxmin} is not solved exactly but approximated by the bottom-up greedy procedure in Algorithm~\ref{alg:form}, inspired by~\cite{DBLP:conf/amcc/FeleMC15}.
Specifically, all the pairwise merges are ranked by resilience first and then by the cost, and the best merges are selected to form the coalitions.
The novelty of the proposed coalition formation lies in the resilience value~\eqref{eq:mu} used to rank and accept the merges.

\begin{algorithm}[tb]
\caption{Resilient coalition formation at time $t$}\label{alg:form}
\begin{algorithmic}[1]
\If{$t\in\mathcal T\setminus\mathcal T^{\mathrm{out}}$}
  \State $\cA\gets\{\mathcal{M}_1,\dots,\mathcal{M}_N\}$, compute $\mu(a)$ by~\eqref{eq:mu} for each $a\in\cA$
  \For{phase $\in\{\textsf{resilience},\textsf{cost}\}$}
    \Repeat
      \State $\mathcal K\gets$ merges $\tilde{\cC}=a\cup b$ of $a\ne b\in\cA$ with $|\tilde{\cC}|\le\mathcal{C}_{\max}$ and $r(\tilde{\cC})\ge0$
      \State compute $\mu(\tilde{\cC})$ by~\eqref{eq:mu} and $\mu_{\mathrm{lo}}=\min\{\mu(a),\mu(b)\}$ for each $\tilde{\cC}\in\mathcal K$
      \If{phase $=$ \textsf{resilience}}
        \State keep in $\mathcal K$ the merges with $\mu(\tilde{\cC})>\mu_{\mathrm{lo}}+\delta$, sorted by increasing $\mu_{\mathrm{lo}}$, then by decreasing $\mu(\tilde{\cC})$
      \Else
        \State solve~\eqref{eq:P} for $a$, $b$ and $\tilde{\cC}$ of each merge in $\mathcal K$, obtain $\mathrm{val}(\tilde{\cC})$, keep the merges with $\mathrm{val}(\tilde{\cC})>0$ and $\mu(\tilde{\cC})\ge\mu_{\mathrm{lo}}-\delta$, sorted by decreasing $\mathrm{val}(\tilde{\cC})$
      \EndIf
      \State $\Anew\gets$ the merges of $\mathcal K$ taken, skipping a merge if $a$ or $b$ is already in a merge taken
      \State $\cA\gets\Anew\cup\big(\cA\setminus(\cup\Anew)\big)$
    \Until{$\Anew=\emptyset$}
  \EndFor
\EndIf
\If{$t\in\mathcal T^{\mathrm{out}}$}
  \State run Steps 2--13 with~\eqref{eq:O} instead of~\eqref{eq:P}, $V^{\mathrm{out}}$ in place of $V$, $\mu=m$, and without $r(\tilde{\cC})\ge0$
\EndIf
\State solve~\eqref{eq:P}, or~\eqref{eq:O} for the coalitions of $\cA$ not solved yet
\State \Return $\cA$ and the plans of its coalitions
\end{algorithmic}
\end{algorithm}

The set $\cA$ holds the coalitions formed at time $t$, where a single building is a coalition with one member, and $\tilde{\cC}=a\cup b$ with $a,b\in\cA$ is a candidate merge.
$V(\cC)$ is the optimal value of~\eqref{eq:P} for $\cC$, and $\mathrm{val}(\tilde{\cC})=V(a)+V(b)-V(\tilde{\cC})$ is the cost saved by the merge, and $\delta$ is a tolerance.
Since $t$ is fixed in Algorithm~\ref{alg:form}, the argument $t$ of $\mu$ and $r$ is omitted.
At every time step, the formation starts from the single buildings, whose resilience values are computed in Step $2$.
For every coalition in $\cA$ and every candidate merge, the lower bound $\st_{\cC}(t+k)$, $k=0,\dots,K-1$ is first computed from~\eqref{eq:stilde} (Steps $2$, $5$ and $6$).
A merge with $r(\tilde{\cC})<0$ has no feasible plan and is skipped without solving (Step $5$).

In the resilience phase, a merge is kept only if it raises the resilience of the weaker coalition of the two, $\mu_{\mathrm{lo}}=\min\{\mu(a),\mu(b)\}$, by more than $\delta$ (Step $8$).
The kept merges are sorted so that the merge with the least resilient weaker coalition comes first, and for the same weaker coalition the partner that makes the merged coalition the most resilient comes first (Step $8$).
Step $11$ then takes the merges in this order and skips a merge whose coalition $a$ or $b$ is already in a taken merge, so each coalition joins at most one merge per round.
The phase ends when no merge raises the resilience of the weaker coalition by more than $\delta$ (Step $13$).
A building with $\mu=-\infty$ is the weakest, so its merges are considered first.
The cost phase then starts from that structure and merges two coalitions if the merge reduces the cost and does not lower the resilience of the weaker one by more than $\delta$, taking the merges with the highest cost saving first (Steps $10$ and $11$).
In this way, cost is reduced only among structures that keep the resilience obtained in the resilience phase.
\begin{remark}
    \label{re,:infeasible}
    Every coalition formed by a merge has $r\ge0$ (Step $5$) and thus a feasible plan, but Algorithm~\ref{alg:form} does not guarantee that every building with $\mu=-\infty$ finds a feasible partner, since this depends on $\mathcal{C}_{\max}$ and on the data.
    If Problem~\eqref{eq:P} has no solution for a building $a$, set $V(a)=+\infty$, and if $a$ is left without a feasible partner, it keeps its battery idle and trades its net demand with the grid.
\end{remark}

During an outage, Step $15$ repeats Steps $2$--$13$ with Problem~\eqref{eq:O}, and the cost phase compares $V^{\mathrm{out}}$, i.e., the load shed and the generation curtailed.
Since no energy can be bought and~\eqref{eq:O} is always feasible, $r$ is not used in Step $5$, and $\mu=m$, where $\st_{\cC}(t)$ remains the energy needed over the next $\tau$ steps.
Reforming coalitions at every step lets buildings with surplus energy supply different buildings as the outage evolves.

Since $\mu$ depends only on the measurements and forecasts at time $t$ and on the battery parameters, Problem~\eqref{eq:P} is solved only in the cost phase (Step $10$) and for the final coalitions (Step $16$).
An optimization problem for each coalition is solved at most once at time $t$.

\subsection{Closed-loop Operation}\label{sec:closed}

Algorithm~\ref{alg:mpc} runs Algorithm~\ref{alg:form} and applies the scheduling plans in a closed loop at every time step.
\begin{algorithm}[tb]
\caption{Closed-loop operation}\label{alg:mpc}
\begin{algorithmic}[1]
\For{$t=1,\dots,T$}
  \State run Algorithm~\ref{alg:form}, obtain the coalition structure $\{\cC_j(t)\}_{j=1}^{M(t)}$ and optimal scheduling
  \For{every building $i$ in $\cC_j(t)$}
    \State apply $u^c_i(t)$, $u^d_i(t)$, $\Cb_i(t)$ and $\Cs_i(t)$
  \EndFor
  \State obtain the realized cost $\sum_j R(\cC_j(t))(t)$ and the states $\SoC_i(t+1)$
  \State $\SoC_{i,t+1}=\SoC_i(t+1)$
\EndFor
\State \Return $J^{\mathrm{MPC}}_{\mathrm{coal}}=\sum_{t}\sum_{j}R(\cC_j(t))(t)$, the $d^{\mathrm{shed}}_i(t)$
  and $g^{\mathrm{curt}}_i(t)$
\end{algorithmic}
\end{algorithm}
At each step $t$, Algorithm~\ref{alg:form} gives the coalitions and their optimal scheduling plan (Step $2$), each building then applies the first step of its plan, and its stored energy is updated (Step $4$).

Finally, the realized cost $R(\cC_j(t))(t)=\sum_{i\in\cC_j(t)}\Pb(t)\,\Gb_i(t)-\Ps(t)\,\Gs_i(t)$, with the energy actually bought and sold, is calculated (Step $5$) and $\SoC_i(t+1)$ is passed to the next step (Step $6$).
The realized cost $J^{\mathrm{MPC}}_{\mathrm{coal}}$, the total load shed and generation curtailed during the outage are recorded.

\section{SIMULATION RESULTS}
This section presents numerical results to evaluate the performance of the proposed resilient coalitional control, including both proactive scheduling and outage management, in a system of $N=12$ buildings.

\subsection{Simulation Setup}
\begin{figure}[tb]
\vspace{4pt}
\centering
\includegraphics[width=\columnwidth]{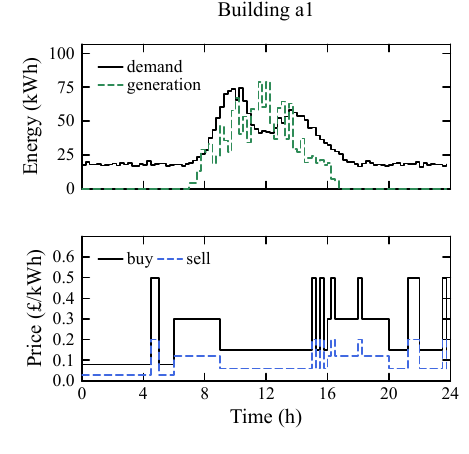}
\caption{Actual demand and generation of building a1, and the prices $\Pb(t)$ and $\Ps(t)$ over the simulated day.}
\label{fig:profiles}
\end{figure}

Our system consists of buildings of six types, denoted from a to f.
Loads and generation units within each building are aggregated, so that each building is represented by a single demand $d_i$, a single generation $g_i$ and a single battery.
The peak demand, the peak generation and the battery parameters of each building are listed in Table~\ref{tab:buildings}, where each battery starts with half of its capacity.
The demand and generation profiles have a sampling time of $T_s=15$ minutes over one day.
Fig.~\ref{fig:profiles} shows, for illustration, the actual demand and generation of building a1 over the simulated day, and the prices $\Pb(t)$ and $\Ps(t)$, which follow a time-of-use tariff with random price peaks.

\begin{table}[tb]
\vspace{4pt}
\caption{Max demand and generation, and battery parameters of the buildings}\label{tab:buildings}
\centering
\setlength{\tabcolsep}{5pt}
\begin{tabular}{@{}lrrrcc@{}}
\hline
Building & Max $d_i$ & Max $g_i$ & $\SoCmax_i$ & $\eta^c_i$ & $\eta^d_i$\\
 & (kWh) & (kWh) & (kWh) & & \\
\hline
a1 & 74.3 & 79.0 & 95 & 0.909 & 0.935\\
a2 & 85.6 & 96.1 & 100 & 0.958 & 0.907\\
b1 & 13.9 & 11.6 & 19 & 0.910 & 0.893\\
b2 & 15.7 & 13.0 & 20 & 0.915 & 0.901\\
b3 & 16.7 & 13.6 & 21 & 0.926 & 0.909\\
c1 & 48.3 & 212.6 & 142.5 & 0.944 & 0.917\\
c2 & 51.9 & 229.8 & 150 & 0.924 & 0.931\\
d1 & 31.1 & 45.5 & 47.5 & 0.922 & 0.878\\
d2 & 34.9 & 51.4 & 50 & 0.911 & 0.899\\
d3 & 39.2 & 55.6 & 52.5 & 0.918 & 0.908\\
e1 & 22.7 & 0.0 & 14.25 & 0.880 & 0.875\\
f1 & 3.0 & 307.5 & 380 & 0.921 & 0.980\\
\hline
\end{tabular}
\end{table}

An upstream grid outage is introduced during $\mathcal T^{\mathrm{out}}=\{63,\dots,83\}$ where the total demand during outage is higher than the generation.
The MPC horizon is $K=8$, the proactive scheduling window is $\tau=4$, the largest coalition size is $\mathcal{C}_{\max}=4$, and the tolerance in Algorithm~\ref{alg:form} is $\delta=10^{-3}$.
During the outage, $c^{\mathrm{shed}}=1$, $c^{\mathrm{curt}}=0.1$ and $\varepsilon=0.05$, which satisfy the condition of Proposition~\ref{prop:nosimO}.
The proposed framework is implemented in Julia~1.12~\cite{bezanson2017julia} with JuMP~\cite{lubin2023jump}, and~\eqref{eq:P} and~\eqref{eq:O} are solved with SCS~\cite{o2016conic}, on a laptop with an Intel Core i7 processor and 32\,GB RAM.\footnote{Code is available at \url{https://github.com/shawn-tao01/Coalition_grid/tree/main}}

\subsection{Results}
The proposed method, denoted RC (Resilient Coalition), forms the coalitions with Algorithm~\ref{alg:form} and applies their plans with Algorithm~\ref{alg:mpc}.
First, the results of RC over the simulated day are presented, then they are compared with decentralized (D0 and D1) and centralized methods (D2).
In the decentralized cases, every building is a coalition of its own.
In D0, each building solves~\eqref{eq:P} without the proactive scheduling constraint~\eqref{eq:floor}.
In D1, each building solves~\eqref{eq:P} with its own reserve,and keeps its battery idle when~\eqref{eq:P} has no solution.
Finally, the centralized case D2 controls all the buildings as a single coalition.

\subsubsection{Resilient coalitional control}
\begin{figure}[tb]
\vspace{4pt}
\centering
\includegraphics[width=\columnwidth]{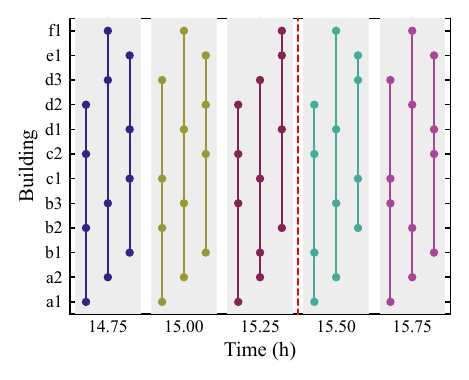}
\caption{Coalitions obtained using the proposed method (RC) between $14.75$\,h and $16$\,h. Each shaded column is one time step labeled with its start time, in which the buildings joined by a vertical line form one coalition. The dashed red line marks the start of the outage.}
\label{fig:coalitions}
\end{figure}

In normal operation, the total stored energy of every coalition is kept above its reserve.
To show how the coalitions share the reserve, we follow buildings c2, a1, b3 and d2, which form one coalition at $15.25$\,h.
Fig.~\ref{fig:coalitions} shows all coalitions of RC between $14.75$\,h and $16$\,h around the start of the outage.
Between $14.75$ and $16$\,h,  a1 is in a coalition with c1 or c2 at every step, and b3 and d2 are in a coalition with c1, c2 or f1, which have the three largest batteries among all buildings.

\begin{figure}[tb]
\centering
\includegraphics[width=\columnwidth]{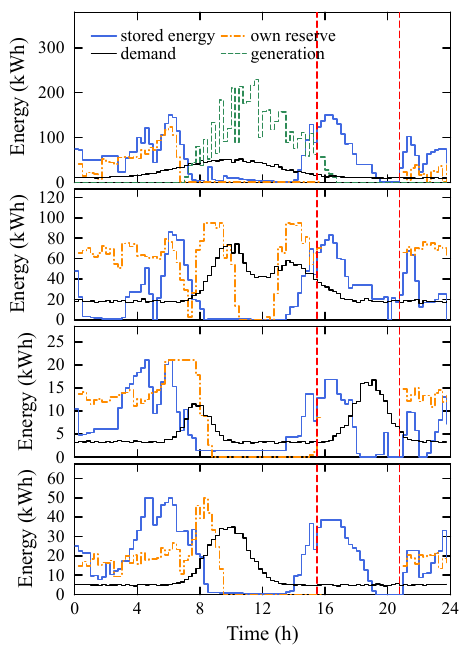}
\caption{Stored energy in RC, own reserve and demand of c2, a1, b3 and d2 from top to bottom, and the generation of c2.}
\label{fig:reserve}
\end{figure}

Fig.~\ref{fig:reserve} shows the stored energy of c2, a1, b3 and d2 for the proposed method RC, together with the reserve~\eqref{eq:stilde} that each would need alone and its demand.
Among them, the generation of c2 exceeds its demand during the day, so its own reserve is zero from $6.75$\,h to the start of the outage.
Before the outage, a1 and b3 hold less than their own reserves at $76\%$ and $39\%$ of the steps, which means their reserve are compensated from other members in their coalitions.
The stored energy in a1, b3 and d2 is almost zero between $9$ and $13$\,h since their batteries are discharged during the price peak from $6$ to $9$\,h.
During the outage, the batteries supply the demand, and the stored energy of the four buildings falls to zero by $20$\,h.
After the outage, a1 and b3 hold less than their own reserves at $11$ of the $12$ steps, while other members of their coalitions keep the reserve.
This shows how coalitions reduce storage requirements on individual buildings.

\begin{figure}[tb]
\centering
\includegraphics[width=\columnwidth]{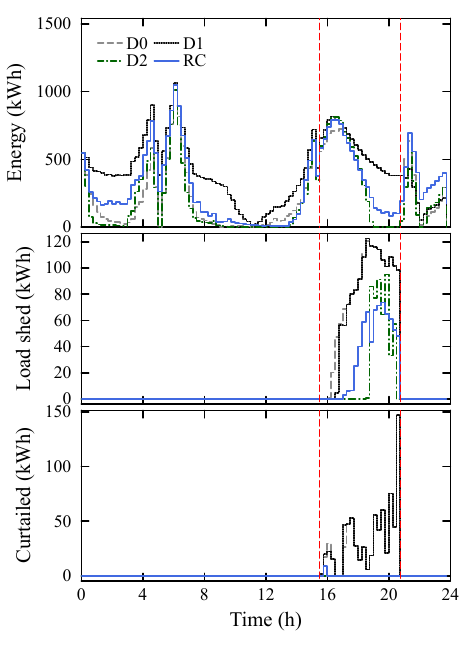}
\caption{Total stored energy, load shed and curtailed generation of the four methods.}
\label{fig:energy_shed}
\end{figure}

Fig.~\ref{fig:energy_shed} shows the total stored energy, the load shed and curtailed generation for methods D0, D1, D2 and RC.
The interval between the dashed red lines is the outage period.
Before the outage, method D1 causes to hold the most energy with $395.5$\,kWh on average, since every building keeps its own reserve, while RC allows to hold a smaller value of $253.7$\,kWh.
During the outage, the generation falls below the demand from $16.5$\,h, and the stored energy starts to decrease.
In D0 and D1, the energy stored by a building cannot supply the others, hence their load shed and the generation curtailment are the largest.
RC sheds none before $17$\,h and $689.1$\,kWh in total, it curtails only $9.3$\,kWh of generation, against $728.3$ and $743.4$\,kWh for D0 and D1.
In the RC plans, there is no simultaneous charging and discharging.

The detailed results of each method are summarized in Table~\ref{tab:results}.
n normal operation, $109$ of the $900$ scheduling problems of D1, that is $12\%$, are infeasible, mostly just after the outage when the batteries are empty, against $17$ and $2$ for RC with $\mathcal{C}_{\max}=2$ and $3$.
D2 and RC with $\mathcal{C}_{\max}=4$ and $5$ have a feasible plan at every step.
The cost of D1 is higher than that of D0, since D0 has no reserve requirements.
RC with $\mathcal{C}_{\max}=4$ costs less than D0 and D1, since within each coalition the surplus of a member supplies the demand of the other members.
It is expected that the centralized method D2 has the lowest cost.
However, it requires sharing the information of all the buildings with a central operator.
Different values of $\mathcal{C}_{\max}$, the maximum number of buildings in each coalition, are also tested.
As $\mathcal{C}_{\max}$ increases, the cost and the load shed decrease, since larger coalitions share the energy of more members.
Load shedding decreases from $968.7$\,kWh for $\mathcal{C}_{\max}=2$ to $547.5$\,kWh for $\mathcal{C}_{\max}=5$, coalitions of two buildings by RC shed $37\%$ less than D1.

\begin{table}[tb]
\caption{Realized cost, load shed, curtailed generation over one day.}\label{tab:results}
\centering
\setlength{\tabcolsep}{4pt}
\begin{tabular}{@{}lrrr@{}}
\hline
Case & Cost & Shed & Curtailed\\
 & (\pounds) & (kWh) & (kWh)\\
\hline
D0 & $-674.2$ & 1618.0 & 728.3\\
D1 & $-537.3$ & 1538.7 & 743.4\\
D2 & $-1031.0$ & 504.9 & 0.0\\
RC, $\mathcal{C}_{\max}=2$ & $-678.3$ & 968.7 & 36.0\\
RC, $\mathcal{C}_{\max}=3$ & $-750.6$ & 849.8 & 22.9\\
RC, $\mathcal{C}_{\max}=4$ & $-836.9$ & 689.1 & 9.3\\
RC, $\mathcal{C}_{\max}=5$ & $-848.2$ & 547.5 & 3.9\\
\hline
\end{tabular}

\end{table}

\section{CONCLUSIONS}
This paper proposed a resilient coalitional control for buildings, designed to mitigate the impact of upstream grid outages through coalition formation, proactive scheduling and outage management.
Simulation results show that, with coalitions of at most four buildings, the proposed method has a feasible plan at every step, and sheds $55\%$ less load and curtails $99\%$ less generation than decentralized control at a lower cost.
Since each coalition can take advantage of a joint energy reserve, before the outage the buildings hold on average $36\%$ less stored energy than operating in a decentralized way with their own reserves only, allowing them more flexibility to improve costs.
Larger coalitions reduce the cost and the load shed.

\bibliographystyle{ieeetr}
\bibliography{bibliography}

@inproceedings{DBLP:conf/amcc/FeleMC15,
  author       = {Filiberto Fele and
                  Jose Maria Maestre and
                  Eduardo F. Camacho},
  title        = {Coalitional control: {A} bottom-up approach},
  booktitle    = {{ACC}},
  pages        = {4074--4079},
  publisher    = {{IEEE}},
  year         = {2015}
}

@article{PANTELI2015259,
title = {Influence of extreme weather and climate change on the resilience of power systems: Impacts and possible mitigation strategies},
journal = {Electric Power Systems Research},
volume = {127},
pages = {259-270},
year = {2015},
issn = {0378-7796},
doi = {https://doi.org/10.1016/j.epsr.2015.06.012},
url = {https://www.sciencedirect.com/science/article/pii/S037877961500187X},
author = {Mathaios Panteli and Pierluigi Mancarella},
}

@article{FILHO2024100158,
title = {Prosumers and sustainable development: An international assessment in the field of renewable energy},
journal = {Sustainable Futures},
volume = {7},
pages = {100158},
year = {2024},
author = {Walter Leal Filho and Laís Viera Trevisan and Amanda Lange Salvia and Janaina Mazutti and Thais Dibbern and Salvador Ruiz {de Maya} and Elvira Ferrer Bernal and João Henrique Paulino Pires Eustachio and Ayyoob Sharifi and María-del-Carmen Alarcón-del-Amo and Iryna Kushnir},
}

@article{Parag2016,
    author = { Parag, Yael and Sovacool, Benjamin K.},
    title = { Electricity market design for the prosumer era},
    journal = {Nature Energy},
    year = {2016},
    volume = {1},
    pages = {16032}
}

@incollection{IPCC2021Ch11,
  author    = {Seneviratne, Sonia I. and Zhang, Xuebin and Adnan, Muhammad and
               Badi, Wafae and Dereczynski, Claudine and Di Luca, Alejandro and
               Ghosh, Subimal and Iskandar, Iskhaq and Kossin, James and
               Lewis, Sophie and Otto, Friederike and Pinto, Izidine and
               Satoh, Masaki and Vicente-Serrano, Sergio M. and Wehner, Michael and
               Zhou, Botao},
  title     = {Weather and Climate Extreme Events in a Changing Climate},
  booktitle = {Climate Change 2021: The Physical Science Basis. Contribution of
               Working Group I to the Sixth Assessment Report of the
               Intergovernmental Panel on Climate Change},
  publisher = {Cambridge University Press},
  address   = {Cambridge, UK and New York, NY, USA},
  pages     = {1513--1766},
  year      = {2021},
  doi       = {10.1017/9781009157896.013},
  url       = {https://www.ipcc.ch/report/ar6/wg1/chapter/chapter-11/}
}

@ARTICLE{7823093,
  author={Fele, Filiberto and Maestre, Jose M. and Camacho, Eduardo F.},
  journal={IEEE Control Systems Magazine}, 
  title={Coalitional Control: Cooperative Game Theory and Control}, 
  year={2017},
  volume={37},
  number={1},
  pages={53-69},
  doi={10.1109/MCS.2016.2621465}}

@article{DBLP:journals/corr/abs-2603-28562,
  author       = {Luke Rickard and
                  Paola Falugi and
                  Eric C. Kerrigan},
  title        = {Coalition Formation with Limited Information Sharing for Local Energy
                  Management},
  journal      = {CoRR},
  volume       = {abs/2603.28562},
  year         = {2026}
}

@article{boyd2011admm,
  author  = {Boyd, Stephen and Parikh, Neal and Chu, Eric and Peleato, Borja and Eckstein, Jonathan},
  title   = {Distributed Optimization and Statistical Learning via the Alternating Direction Method of Multipliers},
  journal = {Foundations and Trends in Machine Learning},
  volume  = {3},
  number  = {1},
  pages   = {1--122},
  year    = {2011},
  doi     = {10.1561/2200000016}
}

@article{nedic2009distributed,
  author  = {Nedi{\'c}, Angelia and Ozdaglar, Asuman},
  title   = {Distributed Subgradient Methods for Multi-Agent Optimization},
  journal = {IEEE Transactions on Automatic Control},
  volume  = {54},
  number  = {1},
  pages   = {48--61},
  year    = {2009},
  doi     = {10.1109/TAC.2008.2009515}
}

@article{HUSSAIN201956,
title = {Microgrids as a resilience resource and strategies used by microgrids for enhancing resilience},
journal = {Applied Energy},
volume = {240},
pages = {56-72},
year = {2019},
issn = {0306-2619},
doi = {https://doi.org/10.1016/j.apenergy.2019.02.055},
url = {https://www.sciencedirect.com/science/article/pii/S0306261919303526},
author = {Akhtar Hussain and Van-Hai Bui and Hak-Man Kim}
}

@article{baldivieso2021coalitional,
  title={Coalitional predictive control: Consensus-based coalition forming with robust regulation},
  author={Baldivieso-Monasterios, Pablo R and Trodden, Paul A},
  journal={Automatica},
  volume={125},
  pages={109380},
  year={2021},
}

@article{sanchez2023robust,
  title={Robust coalitional model predictive control with negotiation of mutual interactions},
  author={S{\'a}nchez-Amores, Ana and Chanfreut, P and Maestre, JM and Camacho, Eduardo F},
  journal={Journal of Process Control},
  volume={123},
  pages={64--75},
  year={2023},
}

@article{oldeWurtel2012mpc,
  title        = {Use of model predictive control and weather forecasts for energy efficient building climate control},
  author       = {Oldewurtel, Frauke and Parisio, Alessandra and Jones, Colin N. and Gyalistras, Dimitrios and Gwerder, Marco and Stauch, Volker and Lehmann, Beat and Morari, Manfred},
  journal      = {Energy and Buildings},
  volume       = {45},
  pages        = {15--27},
  year         = {2012},
}

@article{garifi2020convex,
  title={Convex relaxation of grid-connected energy storage system models with complementarity constraints in DC OPF},
  author={Garifi, Kaitlyn and Baker, Kyri and Christensen, Dane and Touri, Behrouz},
  journal={IEEE Transactions on Smart Grid},
  volume={11},
  number={5},
  pages={4070--4079},
  year={2020},
  publisher={IEEE}
}

@article{zografou2020microgrid,
  title={MicroGrid resilience-oriented scheduling: A robust MISOCP model},
  author={Zografou-Barredo, Natalia-Maria and Patsios, Charalampos and Sarantakos, Ilias and Davison, Peter and Walker, Sara Louise and Taylor, Philip C},
  journal={IEEE Transactions on Smart Grid},
  volume={12},
  number={3},
  pages={1867--1879},
  year={2020},
  publisher={IEEE}
}

@article{vahedipour2020risk,
  title={Risk-based stochastic scheduling of resilient microgrids considering demand response programs},
  author={Vahedipour-Dahraie, Mostafa and Rashidizadeh-Kermani, Homa and Anvari-Moghaddam, Amjad},
  journal={IEEE Systems Journal},
  volume={15},
  number={1},
  pages={971--980},
  year={2020},
  publisher={IEEE}
}

@inproceedings{shimillas2025adaptive,
  title={Adaptive Fault-Aware Reconfiguration of Resilient Distribution Networks Under Emergency Conditions},
  author={Shimillas, Marios and Tziovani, Lysandros and Timotheou, Stelios and Hatziargyriou, Nikos and Panteli, Mathaios},
  booktitle={2025 IEEE Kiel PowerTech},
  pages={1--6},
  year={2025},
  organization={IEEE}
}

@article{bordbari2024networked,
  title={Networked microgrids: A review on configuration, operation, and control strategies},
  author={Bordbari, Mohammad Javad and Nasiri, Fuzhan},
  journal={Energies},
  volume={17},
  number={3},
  pages={715},
  year={2024},
  publisher={MDPI}
}

@article{mansouri2022hierarchical,
  title={A hierarchical scheduling framework for resilience enhancement of decentralized renewable-based microgrids considering proactive actions and mobile units},
  author={Mansouri, Seyed Amir and Nematbakhsh, Emad and Ahmarinejad, Amir and Jordehi, Ahmad Rezaee and Javadi, Mohammad Sadegh and Marzband, Mousa},
  journal={Renewable and Sustainable Energy Reviews},
  volume={168},
  pages={112854},
  year={2022},
  publisher={Elsevier}
}

@article{modarresi2022coalitional,
  title={Coalitional game theory approach for multi-microgrid energy systems considering service charge and power losses},
  author={Modarresi, Javad},
  journal={Sustainable Energy, Grids and Networks},
  volume={31},
  pages={100720},
  year={2022},
  publisher={Elsevier}
}

@article{gautam2023graph,
  title={A graph theory and coalitional game theory-based pre-positioning of movable energy resources for enhanced distribution system resilience},
  author={Gautam, Mukesh and Benidris, Mohammed},
  journal={Sustainable Energy, Grids and Networks},
  volume={35},
  pages={101095},
  year={2023},
  publisher={Elsevier}
}

@article{bezanson2017julia,
  title={Julia: A fresh approach to numerical computing},
  author={Bezanson, Jeff and Edelman, Alan and Karpinski, Stefan and Shah, Viral B},
  journal={SIAM review},
  volume={59},
  number={1},
  pages={65--98},
  year={2017},
  publisher={SIAM}
}

@article{lubin2023jump,
  title={JuMP 1.0: Recent improvements to a modeling language for mathematical optimization},
  author={Lubin, Miles and Dowson, Oscar and Garcia, Joaquim Dias and Huchette, Joey and Legat, Beno{\^\i}t and Vielma, Juan Pablo},
  journal={Mathematical Programming Computation},
  volume={15},
  number={3},
  pages={581--589},
  year={2023},
  publisher={Springer}
}

@article{o2016conic,
  title={Conic optimization via operator splitting and homogeneous self-dual embedding},
  author={O’donoghue, Brendan and Chu, Eric and Parikh, Neal and Boyd, Stephen},
  journal={Journal of Optimization Theory and Applications},
  volume={169},
  number={3},
  pages={1042--1068},
  year={2016},
  publisher={Springer}
}

@article{casagrande2022resilient,
  title={Resilient microgrid energy management algorithm based on distributed optimization},
  author={Casagrande, Vittorio and Prodan, Ionela and Spurgeon, Sarah K and Boem, Francesca},
  journal={IEEE Systems Journal},
  volume={17},
  number={2},
  pages={2791--2802},
  year={2022},
  publisher={IEEE}
}

@article{casagrande2024learning,
  title={Learning-based MPC with uncertainty estimation for resilient microgrid energy management},
  author={Casagrande, Vittorio and Ferianc, Martin and Rodrigues, Miguel and Boem, Francesca},
  journal={IFAC-PapersOnLine},
  volume={58},
  number={4},
  pages={556--561},
  year={2024},
  publisher={Elsevier}
}

@inproceedings{tao2025resilient,
  title={Resilient Microgrid Energy Management Under Grid Faults: A Convex Optimization Approach},
  author={Tao, Haochen and Tziovani, Lysandros and Casagrande, Vittorio and Timotheou, Stelios and Boem, Francesca},
  booktitle={2025 6th International Conference on Control and Fault-Tolerant Systems (SysTol)},
  pages={89--94},
  year={2025},
  organization={IEEE}
}

@article{casagrande2024online,
  title={Online end-to-end learning-based predictive control for microgrid energy management},
  author={Casagrande, Vittorio and Ferianc, Martin and Rodrigues, Miguel RD and Boem, Francesca},
  journal={IEEE Transactions on Control Systems Technology},
  volume={33},
  number={2},
  pages={463--478},
  year={2024},
  publisher={IEEE}
}




\section*{APPENDIX}
\subsection{Proof of Proposition~\ref{prop:nosimO}}
The proof of i) follows~\cite{garifi2020convex}, where any simultaneous charging and discharging losses could simply be sold to improve the solution.
Hence, any solution with simultaneous charging and discharging is necessarily suboptimal.

For ii), let $x_{i,k}$ denote $x_i(t+k)$.
Since~\eqref{eq:O} is a linear program, every optimal solution satisfies the KKT conditions.
Let $\nu_{i,k}$, $\pi_{i,k}$ and $\xi_k$ be the Lagrange multipliers of~\eqref{eq:Opower},~\eqref{eq:dyn} and~\eqref{eq:coal}, let $\lambda^{c\pm}_{i,k}$, $\lambda^{d\pm}_{i,k}$, $\kappa^{\pm}_{i,k}$ and $\sigma^{\pm}_{i,k}\ge0$ be those of the upper and lower bounds of $u^c_{i,k}$, $u^d_{i,k}$, $g^{\mathrm{curt}}_{i,k}$ and $d^{\mathrm{shed}}_{i,k}$ respectively, and let $\rho^b_{i,k},\rho^s_{i,k}\ge0$ be multipliers of $\Cb_{i,k},\Cs_{i,k}\ge0$.
With~\eqref{eq:Opower} and~\eqref{eq:dyn} written with all terms on the left-hand side, the stationarity conditions with respect to $u^c_{i,k}$, $u^d_{i,k}$, $g^{\mathrm{curt}}_{i,k}$ and $d^{\mathrm{shed}}_{i,k}$ are
\begin{align}
&\varepsilon+\nu_{i,k}-\eta^c_i\pi_{i,k}+\lambda^{c+}_{i,k}-\lambda^{c-}_{i,k}=0, \label{eq:kkt_c}\\
&-\nu_{i,k}+\pi_{i,k}/\eta^d_i+\lambda^{d+}_{i,k}-\lambda^{d-}_{i,k}=0, \label{eq:kkt_d}\\
&c^{\mathrm{curt}}+\nu_{i,k}-\kappa^-_{i,k}+\kappa^+_{i,k}=0, \label{eq:kkt_curt}\\
&c^{\mathrm{shed}}-\nu_{i,k}-\sigma^-_{i,k}+\sigma^+_{i,k}=0, \label{eq:kkt_shed}
\end{align}
and those with respect to $\Cb_{i,k}$ and $\Cs_{i,k}$ give $\nu_{i,k}=\xi_k-\rho^b_{i,k}\le\xi_k$ and $\nu_{i,k}=\xi_k+\rho^s_{i,k}\ge\xi_k$, so $\nu_{i,k}=\xi_k$ for all $i\in\cC_j$.

By contradiction, let $u^c_{i,k}>0$ and $u^d_{i,k}>0$ in an optimal solution of~\eqref{eq:O}, at a step $k$ at which a member $h\in\cC_j$ sheds load or does not curtail all its generation.
Complementary slackness gives $\lambda^{c-}_{i,k}=\lambda^{d-}_{i,k}=0$.
Multiplying~\eqref{eq:kkt_d} by $\eta^c_i\eta^d_i$ and adding it to~\eqref{eq:kkt_c} cancels $\pi_{i,k}$, and substituting $\nu_{i,k}=\xi_k$ gives
\begin{equation*}
\varepsilon+\xi_k(1-\eta^c_i\eta^d_i)=-\lambda^{c+}_{i,k}-\eta^c_i\eta^d_i\lambda^{d+}_{i,k}\le0,
\end{equation*}
so with $\varepsilon>c^{\mathrm{curt}}(1-\eta^c_i\eta^d_i)$,
\begin{equation}\label{eq:xi_bound}
\xi_k\le-\varepsilon/(1-\eta^c_i\eta^d_i)<-c^{\mathrm{curt}}.
\end{equation}
If $h$ does not curtail all its generation, then $\kappa^+_{h,k}=0$, and substituting it and $\nu_{h,k}=\xi_k$ into~\eqref{eq:kkt_curt} gives $\xi_k=-c^{\mathrm{curt}}+\kappa^-_{h,k}\ge-c^{\mathrm{curt}}$.
If $h$ sheds load, then $\sigma^-_{h,k}=0$, and~\eqref{eq:kkt_shed} gives $\xi_k=c^{\mathrm{shed}}+\sigma^+_{h,k}>0$.
Both contradict~\eqref{eq:xi_bound}. \hfill$\blacksquare$

\end{document}